\documentclass[fleqn,10pt]{wlscirep}
\usepackage[utf8]{inputenc}
\usepackage[T1]{fontenc}
\usepackage{bm}
\usepackage{cite}
\usepackage{float}             
\usepackage{amsmath,amssymb}   
\usepackage{amsfonts}
\usepackage{graphicx}          
\usepackage{graphics}
\usepackage{hyperref}

\title{Ionization energy of atoms in photonic crystals}

\author[1,2*]{Renat Kh. Gainutdinov}
\author[1]{Adel I. Garifullin}
\author[1,2]{Marat A. Khamadeev}
\author[2,1]{Myakzyum Kh. Salakhov}
\affil[1]{Institute of Physics, Kazan Federal University, Kazan, 420008, Russian Federation}
\affil[2]{Tatarstan Academy of Sciences, Kazan, 420013, Russian Federation}

\affil[*]{Renat.Gainutdinov@kpfu.ru}

\keywords{Quantum electrodynamics and energy levels of atoms in photonic crystals, the electron mass in photonic crystals, Ionization energies}

\begin{abstract}
The periodic changes in physical and chemical properties of the chemical elements is caused by the periodic change of the ionization energies. The ionization energy of each element is constant and this manifests itself in the periodic table. However, we show that the
ionization energies can be dramatically changed, when atoms are placed in a photonic crystal consisting of materials with a highly tunable
refractive index and voids. The tunability of these materials gives rise to the tunability of the ionization energies over a wide range. This allows one to come beyond the limitations put on by the periodic table on physical and chemical processes, and can open up new horizons in synthesizing exceptional chemical compounds that could be used in pharmaceutical and other medical-related activities.
\end{abstract}

\begin{document}

\flushbottom
\maketitle

The United Nations has designated 2019 as the international year of the periodic table of chemical elements being one of the most important
tools in science \cite{IYPT2019}. The periodicity has been explained in consistent way by quantum theory of the atom, and this explanation is
founded on the concept of electronic shells and subshells. The periodic changes in physical and chemical properties of the elements is caused
by the periodic change of the ionization energies being the minimal energy required to remove the outermost electron from an atom with the
atomic number \textit{Z}. The ionization energy of each element is constant and this manifests itself in the periodic table. However, as we
show in this paper the ionization energies can be dramatically changed, when atoms are placed in air voids of a photonic crystal. The control
of the values of the ionization energies that play a key role in chemical reactions can allow one to synthesize exceptional chemical compounds that could be used in pharmaceutical and other medical-related activities.

Photonic crystals (PCs) consist of spatially varying structures that are periodic on the scale of the wavelength of the incident light
\cite{Yablonovitch1987,John1987,Lopez2003,Joannopoulos2008,Soukoulis2001photonic,Von2013bottom}. Photonic crystals have many potential
technological applications
\cite{John1990quantum,John1991,Quang1997coherent,Zhu1997quantum,Bay1997atom,Busch2000radiating,Aguirre2010tunable,Huang2011dirac,Fenzl2014photonic,Goban2014atom,Segal2015controlling,Gainutdinov2012}.
Many interesting novel effects on optical behavior of atoms and quantum dots in photonic crystals such as strong emitter-photon coupling, coherent control of spontaneous emission, modification of Lamb shift, $\it{etc}$
have been predicted and observed \cite{John1990quantum,John1991,Quang1997coherent,Zhu2012highly,Zhu2012inhibited,Liu2010observation,Roy2010coherent,Yoshie2004vacuum,Vats2002theory,Li2001quantum,Wang2004giant,Wang2005spontaneous}. These applications and effects are mainly based on the photonic band gap effect. However, the band gap is not only an effect that follows from the periodic changing of the refractive index in PCs. In Ref.\cite{Gainutdinov2012} it has been shown that the modification of the interaction of an electron placed in air voids of a PC with its own radiation field results in the change of its electromagnetic mass. This effect can open up new possibilities for applying PCs such as the realization of quantum interference, creation of new light sources with line spectrum, new $He$-$Ne$ like lasers, \textit{etc} \cite{Gainutdinov2018}.

\subsection*{Anisotropy of the electron mass in the photonic crystal medium.}

The change in ionization energies of atoms is one of the consequences of the effect of the change of the electromagnetic mass of an electron
placed in the PC medium. Most explicitly this effect manifests itself in the case when a PC consists of regularly arranged voids in a
dielectric with a highly tunable refractive index. The typical size of these voids is $10^3$ times larger than the Bohr radius. This means
that an atom in a void behaves itself as a free one. Only the interaction of the atom with its own radiation field is modified in the PC
medium.
This self-interaction consists of the interaction of each atomic electron with its own radiation field and the self-interaction of the atom
involving the Coulomb interaction of the atomic electrons with the nucleus. In the case of the atom in the free space, the first interaction
process generates the electromagnetic mass of the electron while the second one gives rise to the Lamb shifts of the atomic energy levels
\cite{Weinberg1995}.

The physical mass $m_e$ of the electron is a sum of its bare $m_0$ and electromagnetic $m_{em}$ masses
\begin{equation}\label{eqn1}
m_e = m_0 + m_{em}.
\end{equation}
However, this self-energy cannot be calculated because of the non-renormalizable ultraviolet divergences. The problem is solved by
including the electromagnetic mass into the physical mass $m_e$ being the only observable, and for these reasons, it must not be calculated.
This is a part of the renormalization procedure being the cornerstone of quantum electrodynamics. The above may be a reason why for a long
time since the pioneering works of Yablonovitch \cite{Yablonovitch1987} and John \cite{John1987} no attention had been paid to the fact
that the modification of the electromagnetic interaction in the PC medium leads to the correction $\delta m_{pc}$ to the electron self-energy
that cannot be hidden in the electron physical mass. In contrast to the electromagnetic mass of an electron in vacuum the PC correction to
this mass is an observable. Actually, in this case we deal with a fundamental quantum electrodynamic effect that manifests itself only in
artificial mediums like the photonic crystal one. For the first time the electromagnetic mass comes to play in describing physical processes.
The mass correction is anisotropic and depends on the electron states. The photonic crystal mass correction is an observable and is
described by an operator. The interaction described by this operator can be very strong and can be comparable with the interaction of the
valence electrons with the atomic nucleus, and significantly affects on ionization energies of atoms in the photonic crystal medium. As we
show in this paper the absolute value of the correction to the ionization energies increases quadratically with refractive index of the
material host of the PC medium. Recently a great progress has been achieved in the design of metamaterials with unnaturally highly tunable
refractive indices \cite{Lee2015,Chung2016,Kim2016,Kim2018}. This makes the effect under study also tunable, and opens the door to novel
technologies in physics, chemistry and pharmacology.

The electromagnetic mass of an electron in vacuum is generated by its interaction with its own electromagnetic field. One of the processes
that gives the contributions to the electron self-energy is the emission and then absorption of a photon. In the Coulomb gauge this process is
separated into the two parts. The process of the interaction of the electron with its own Coulomb field which leads to the change in the
electron mass, and the process of its interaction with its own transverse field gives rise to the correction of its kinetic energy (here and below, we use the natural system of units in which $\hbar = c = 1$):
\begin{equation}\label{eqn2}
\Delta {E_e}\left( {\bf{p}} \right) = \Delta {m_{em}} + \frac{{\Delta m_{em}^{'}}}{{2m_e^2}}{{\bf{p}}^2} + O{\left( {\left| {\bf{p}} \right|} \right)^3} + O\left( {\frac{{\Delta {m_{em}}}}{{{m_e}}}} \right),
\end{equation}
where
\begin{equation}\label{eqn3}
\Delta m_{em}^{'} = \frac{\alpha }{{{\pi ^2}}}\int\limits_{0}^{k_0} {\frac{{{d^3}{\bf{k}}}}{{2{{\bf{k}}^2}}}} \sum\limits_{\lambda  = 1}^2 {{{\left| {{{\bf{I}}_{\bf{p}}} \cdot {{\bm{\varepsilon }}_\lambda }({\bf{k}})} \right|}^2}} = \frac{{{4\alpha }}}{{{3{{\pi }}}}}{k_0}
\end{equation}
with ${{\bf{I}}_{\bf{{p}}}} = \frac{{\bf{p}}}{\left| {\bf{p}} \right|}$  being the direction of the electron momentum, $\bm{\varepsilon}
_\lambda  ({\bf{k}})$ is the unit vector of the field polarization ($\lambda$) in free space, and $\alpha$ is the fine-structure constant.
The correction $\Delta m_{em}^{'}$ to the electron mass is determined by a divergent integral, unless a cutoff $k_0$ is introduced \cite{Cohen1998atom}. This electron mass
correction $\Delta m_{em}^{'}$ that appears in the expression for the correction to the kinetic energy of the electron must coincide with
the mass correction $\Delta m_{em}$ being the result of the interaction of the electron with its own Coulomb field. This follows from the
relativistic energy-momentum relationship for an electron in vacuum
\begin{equation}
E^2 - {\textbf{p}}^2 = m_e^2 \notag \\
\end{equation}
with $E^2 - {\textbf{p}}^2$ being a Lorentz invariant.
And really, the mass correction $\Delta m_{em}$ describing the one-photon Coulomb self-energy with the same cutoff is given by equation \cite{Cohen1998atom}
\begin{equation}\label{eqn4}
\Delta m_{em} = \frac{{{4\alpha }}}{{{3{{\pi }}}}}{k_0}.
\end{equation}
Thus $\Delta m_{em}$ defined in Eq.~(\ref{eqn4}) is a correction to the electron mass from the self-energy processes associated with the emission
and absorption of virtual photons. In the case where the electron interacts with the PC electromagnetic field the photons are replaced with
the Bloch photons (see Methods for details). In the PC medium vacuum becomes anisotropic, and $E^2 - \textbf{p}^2$ is an invariant only under
the Lorentz boost with the velocity $\bf{v}$ directed along the direction ${{\bf{I}}_{\bf{{p}}}}$ of the electron momentum. The "invariant"
depends on this direction
\begin{equation}\label{eqn5}
E^2-{\textbf{p}}^2 = m_{pc}^2({\bf{I}}_{\bf{{p}}}),
\end{equation}
where $m_{pc} = m_e + \delta m_{pc}({\bf{I}}_{\bf{{p}}})$. The electron mass correction $\Delta m_{em}$ defined in Eq.~(\ref{eqn4}) is finite only owing
to the cutoff $k_0$, and there are no renormalization procedure that could be used to remove such cutoffs. This means that the ultraviolet divergences
in equations describing the electromagnetic mass of the electron are not renormalizable, and only the way to solve the problem is to include the electromagnetic
mass into the observable mass $m_e$. In the case where we deal with the electron interacting with the PC vacuum, only the part of the electromagnetic mass
equals to the electromagnetic mass $\Delta m_{em}$ can be hidden in the observable electron mass $m_e$. In this way we arrive at the observable correction to the
electron mass
\begin{equation}\label{eqn6}
\delta m_{pc} = \Delta m_{em}^{pc} - \Delta m_{em} = \frac{\alpha }{{\pi ^2 }}\left[ {\sum\limits_{n}  \int\limits_{FBZ}  \,\frac{{d^3 {\bf{k}} }}{{\omega _{{\bf{k}}n}^2
}}\sum\limits_{\bf{G}} {\left| {{{\bf{I}}_{\bf{p}}}\cdot {\bf{E}}_{{\bf{k}}n} ({\bf{G}})} \right|^2 } } -  {\int {\frac{{d^3 {\bf{k}}}}{{2{\bf{k}}^2 }}} \sum\limits_{\lambda  = 1}^2  \,\mathop {\left| { {{\bf{I}}_{\bf{p}}}\cdot
 \bm{\varepsilon} _\lambda  ({\bf{k}})} \right|}\nolimits^2 } \right]
\end{equation}
(see Eq.~(\ref{eqn18}) in Methods).
It is important that in contrast to the electron electromagnetic mass $m_{em}$ the PC correction to the electron mass is free from ultraviolet divergences (see Supplementary Section 3 for details),
because there is a natural cutoff ${\Lambda _N}$ above which the refractive index $n(\omega)$ of the dielectric host tends to unity, and, as a consequence, the difference between the PC and vacuum contributions to $\delta m_{pc}$ becomes negligible. This allows one to estimate $\delta m_{pc}$ as
\begin{equation}\label{eqn61}
\left\langle {\delta {m_{pc}}} \right\rangle \sim \frac{\alpha}{\pi} {\Lambda _N}{\left\langle {n\left( \omega  \right)} \right\rangle ^2},
\end{equation}
where $\left\langle {n\left( \omega  \right)} \right\rangle$ is the mean value of the refractive index. In contrast, the Lamb shifts of the energy levels of the atomic hydrogen embedded into the voids of PCs are of
order $\frac{{4{\alpha ^5}}}{{3\pi {n^3}}}\left( {\ln \frac{{{m_e}}}{{2\bar E}} + \frac{{11}}{{24}} - \frac{1}{5}} \right){m_e}$, where $\bar E$ = 8,9${\alpha ^2}{m_e}$,
$n$ is the principal quantum number, although for some states of an atom, the Lamb shift is not more than two orders greater \cite{Wang2004giant}.
Even being enhanced by 2 order the Lamb shift in the atomic hydrogen is of order $4 \cdot {10^{ - 4}}$ $eV$.
In the case of high refractive index materials used in our work the cutoff is of order 35 $eV$, and the mean value of the refractive index
$\left\langle {n\left( \omega  \right)} \right\rangle$ is about 8, the mass correction $\left\langle {\delta {m_{pc}}} \right\rangle $
describing the atomic energy shifts is about 5 $eV$. For this reason, below we will neglect the contributions from the Lamb shifts to the atomic
transition frequencies.

The operator associated with the observable $\delta m_{pc}$ is as follows \cite{Gainutdinov2012}
\begin{equation}\label{eqn7}
\widehat {\delta {m}}_{pc}\left( \widehat {\bf{I}}_{\bf{p}} \right)  = \frac{\alpha }{{\pi ^2 }}\left[ {\sum\limits_{n}  \int\limits_{FBZ}  \,\frac{{d^3 {\bf{k}} }}{{\omega
_{{\bf{k}}n}^2 }}\sum\limits_{\bf{G}} {\left| {{\widehat {\bf{I}}_{\bf{p}}}\cdot {\bf{E}}_{{\bf{k}}n} ({\bf{G}})} \right|^2 } } -  {\int {\frac{{d^3 {\bf{k}}}}{{2{\bf{k}}^2 }}} \sum\limits_{\lambda  = 1}^2  \,\mathop {\left| { {\widehat {\bf{I}}_{\bf{p}}}\cdot
 \bm{\varepsilon} _\lambda  ({\bf{k}})} \right|}\nolimits^2 } \right]. \\
\end{equation}
The expression for $\widehat {\delta {m}}_{pc}$ is independent of the atomic potential or the periodic potentials in solids, and are the same for a free electron.
This follows from one of the primary principles of physics is the principle of superposition of probability amplitudes: \textit{"The probability amplitude of an event which can happen in different ways
is a sum of the probabilities amplitudes for each of this way"}, which has been formulated by R. Feynman in Ref.~\cite{Feynman1948}. In the problem under study, an alternative way is the process in which the electron
interacts only with its own radiation field, and just this process gives rise to the change in the electron rest mass. Other alternatives associated with involving of the electron interactions with
potentials give not contributions to the electron mass. In the case of atomic electrons, the involving of the Coulomb interaction into the self-energy process gives rise to the Lamb shifts of atomic states.
Hamiltonians of atoms in the PC medium must be completed by the operators $\widehat {\delta {m}}_{pc}$ for each electron. For example, the Hamiltonian of the atomic hydrogen takes the form
\begin{equation}\label{eqn10}
H_{pc} = \widehat {\delta {m}}_{pc} + {H},
\end{equation}
where $H$ is the Hamiltonian of the atomic hydrogen in free space containing the rest mass part which is not usually considered. The atomic states and energies thus are determined by equation
\begin{equation}\label{eqn11}
H_{pc}\left| {{\Psi _{i,pc}}} \right\rangle  = {E_{i,pc}}\left| {{\Psi _{i,pc}}} \right\rangle.
\end{equation}
This equation can be solved perturbatively expanding the solution in powers of $\widehat {\delta {m}}_{pc}$. At leading order of the solution we get $\left| {{\Psi _{i,pc}^{(1)}}} \right\rangle  = \left| {{\Psi _{i}}} \right\rangle$ and $E_i^{(1)} = \left\langle {{\Psi _i}}\right|\widehat {\delta {m}}_{pc}\left| {{\Psi _i}} \right\rangle  + {E_i}$, where ${E_i}$ is the energy of the state $\left| {{\Psi _i}}\right\rangle$ of the atom in the free space.
It should be noted that the correction $E^{(1)}_i$ - $E_i$ depends only on orbital $l$ and magnetic $m_l$ quantum numbers but not on the electron coordinate,
$\left\langle {{\Psi}} \right|\widehat {\delta m}_{pc}\left| {{\Psi}} \right\rangle = \left\langle {{l,m_l}}\right|\widehat {\delta m}_{pc}\left| {{l,m_l}} \right\rangle$. For the case of hydrogen atom and the alkali metals $l$ = 0, $m_l = 0$.
Hamiltonian of an $\textit{N}$-electron atom placed in PC medium is described in the similar way:
\begin{equation}\label{eqn12}
H_{pc}^N = \sum\limits_{i = 1}^N {\widehat {\delta m}_{pc}^{(i)}}  + H_{}^N,
\end{equation}
where $H^N$ also contains the rest masses of each electron. Since these parts are always constant they do not appear in the energy of the
transition between states of the atom in vacuum. But in PC there are corrections to these energies that could significantly modify
familiar optical spectra of atoms. Transitions between states of the $N$-electron atom are accompanied by changes of the configurations of
electrons in the subshells and PC medium corrections to its energies could have complex structure.

\subsection*{Ionization energy of atoms in one-dimensional photonic crystals.}

\begin{figure}[h]
\centering
\includegraphics[width=0.7\linewidth]{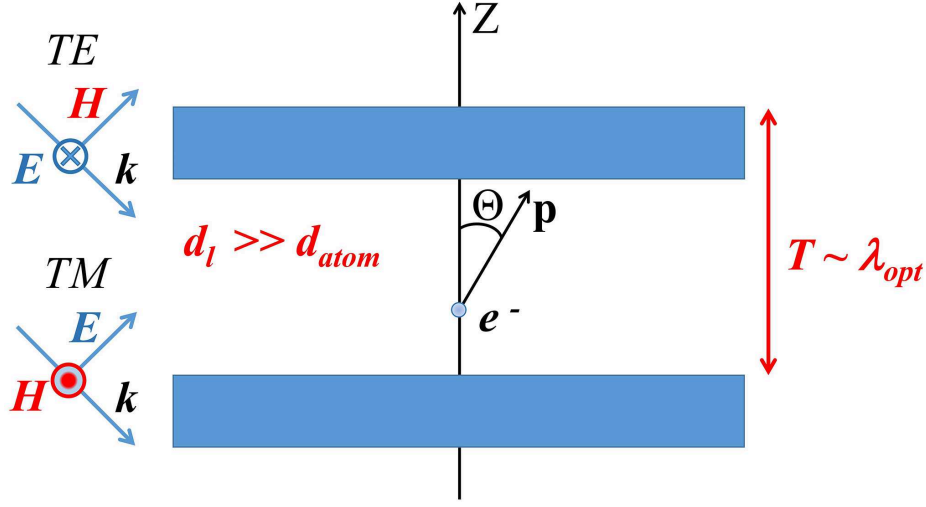}
\caption{\label{fig1} Free electron placed in an air void of the one-dimensional PC having cylindrical symmetry. Electron propagates at an
$\Theta$-angle to the $\textit{Z}$\;-\;axis of crystal. The size of the vacuum's layer of PC is $10^3$ times larger than the average size of
atom.}
\end{figure}

Let us consider one-dimensional PCs (Fig.~\ref{fig1}) because they are most important for applications of the effect under study. For the case of one-dimensional
PC with the selected $\textit{Z}$\;-\;axis of crystal the coefficients ${\bf{E}}_{{\bf{k}}n} ({\bf{G}})$ have the
polarization structure
\begin{equation}\label{eqn8}
{{\bf{E}}_{{\bf{k}}n}}(\textbf{G}) = \sum\limits_{\lambda = 1}^2 {{E}_{{\bf{k}}n\lambda}({G})}{{\bm{\varepsilon }}_{\lambda}
(\bf{k_G})},
\end{equation}
where ${\bm{\varepsilon}}_{{1}}(\bf{k_G}) $ and ${\bm{\varepsilon}}_{{2}}(\bf{k_G}) $ are unit vectors of the TE (transverse-electric) and TM
(transverse-magnetic) polarization, correspondingly, ${\bf{k_G}} = {\bf{k}} + G{{\bf{e}}_z}$.

\noindent Thus, for 1D PC the operator of the self-energy correction Eq.~(\ref{eqn7}) can be rewritten as (see Supplementary Section 2 for details)
\begin{equation}\label{eqn9}
\widehat {\delta {m}}_{pc} = A + \left( {{{\widehat {\bf{I}}}_{\bf{p}}} \cdot {{\widehat {\bf{I}}}_{pc}}} \right)^{2}B,
\end{equation}
where ${{{\widehat {\bf{I}}}_{pc}}}$ is the unit vector of the 1D PC crystal axis that coincides with vector ${\bf{e}}_z$ and
\begin{equation}
\begin{split}
A = \frac{{{\alpha }}}{{{\pi }}}\sum\limits_{n,\;G}^{} {\int {{k_\rho }d{k_\rho }} \int\limits_{FBZ}^{} {d{k_z}\left( {\frac{{{{\left|
{E_{{\bf{k}}n1}^{}\left( G \right)} \right|}^2}}}{{{{\omega }}_{{\bf{k}}n1}^2}}\frac{{k_{Gz}^2}}{{k_\rho ^2 +k_{Gz}^2}}}
 + \frac{{{{\left| {E_{{\bf{k}}n2}^{}\left( G \right)} \right|}^2}}}{{{{\omega }}_{{\bf{k}}n2}^2}} \right)}}-\frac{{4{{\alpha
}}}}{{3{{\pi }}}}{\int {dk}},\notag \\
\end{split}
\end{equation}
\begin{equation}
\begin{split}
B = \frac{{{\alpha }}}{{{\pi }}}\sum\limits_{n,\;G}^{} {\int {{k_\rho }d{k_\rho }} \int\limits_{FBZ}^{} {d{k_z}\left( {\frac{{{{\left|
{E_{{\bf{k}}n1}^{}\left( G \right)} \right|}^2}}}{{{{\omega }}_{{\bf{k}}n1}^2}}\frac{{2k_\rho ^2 - k_{Gz}^2}}{{k_\rho ^2 + k_{Gz}^2}} - \frac{{{{\left| {E_{{\bf{k}}n2}^{}\left( G \right)} \right|}^2}}}{{{{\omega }}_{{\bf{k}}n2}^2}}} \right)}}.\notag \\
\end{split}
\end{equation}
Here $\omega _{{\bf{k}}n1}$ and $\omega _{{\bf{k}}n2}$ are dispersion relations for TE and TM Bloch modes satisfying transcendent equation
\cite{Skorobogatiy2009} (See Methods for details of dispersion relations).

Let us consider the ionization process for the hydrogen atom and the alkali metals in which the transition is defined only by one valence electron. In the case the lower state is the ground state of an atom ($S$-state) and the upper state is the free state.
The difference between energies of these states determines the binding energy of an electron $E_{bind}$. It should be noted that in the
case of PC medium $E_{bind}^{pc}$ depends on the direction of a free electron momentum $\bf{p}$ that can impact on configurations of bonds in molecules. But primarily the PC medium corrections modify the ionization energy of atoms which is defined as the minimum energy necessary to remove an electron. According to this definition, the correction to the ionization energy of an atom is determined by equation:
\begin{equation}\label{eqn13}
\delta E_{ion}^{pc} = \delta m_{pc}^{min} - {\delta m_{pc}^{{l,m}}},
\end{equation}
where $\delta m_{pc}^{min}$ is the smallest correction determining by Eq. (\ref{eqn9}), and
${\delta m_{pc}^{{l,m}}} = \left\langle {{\Psi}} \right|\widehat {\delta m}_{pc}\left| {{\Psi}} \right\rangle = \left\langle {{l,m_l}}
\right|\widehat {\delta m}_{pc}\left| {{l,m_l}} \right\rangle$. For the case of hydrogen atom and the alkali metals $l$ = 0, $m_l$ = 0.

\begin{figure}[h]
\centering
\includegraphics[width=0.7\linewidth]{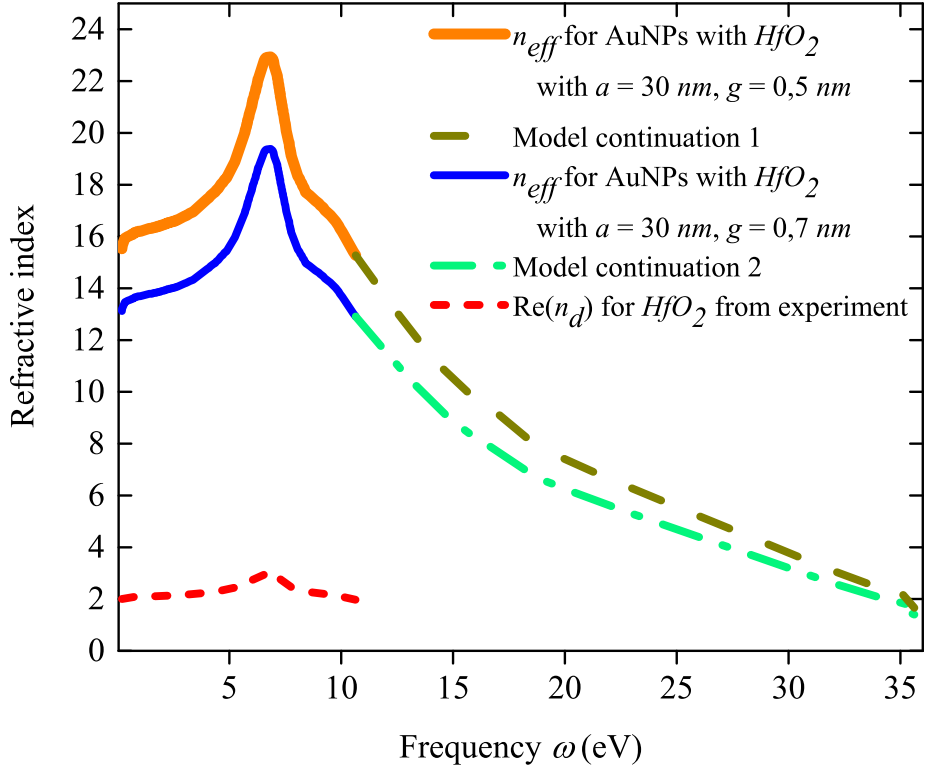}
\caption{\label{fig2} The spectral dependencies $n_{eff}(\omega)$ for metamaterial consisting of AuNPs ensemble coated with ${HfO_2}$ (thick
orange and blue solid lines) extracted from the experiment ($n_d(\omega)$ of the ${HfO_2}$) (red dashed line) \cite{Franta2015}. These curves
for the same region of frequencies as the experimental one have been plotted using equation $n_{eff} = [(a/g)\varepsilon_d]^{1/2}$
\cite{Chung2016} with $\textit{a}$ = 30 $nm$, $g$ = 0.5 $nm$ (thick orange solid line), and $\textit{a}$ = 30 $nm$, $g$ = 0.7 $nm$ (blue solid
line). The rest parts of these curves have been chosen to provide the fact that at the high frequencies $n_{eff}$$\rightarrow$1
(brown dashed line and green dash-dotted line).}
\end{figure}

\begin{figure}[h]
\centering
\includegraphics[width=0.7\linewidth]{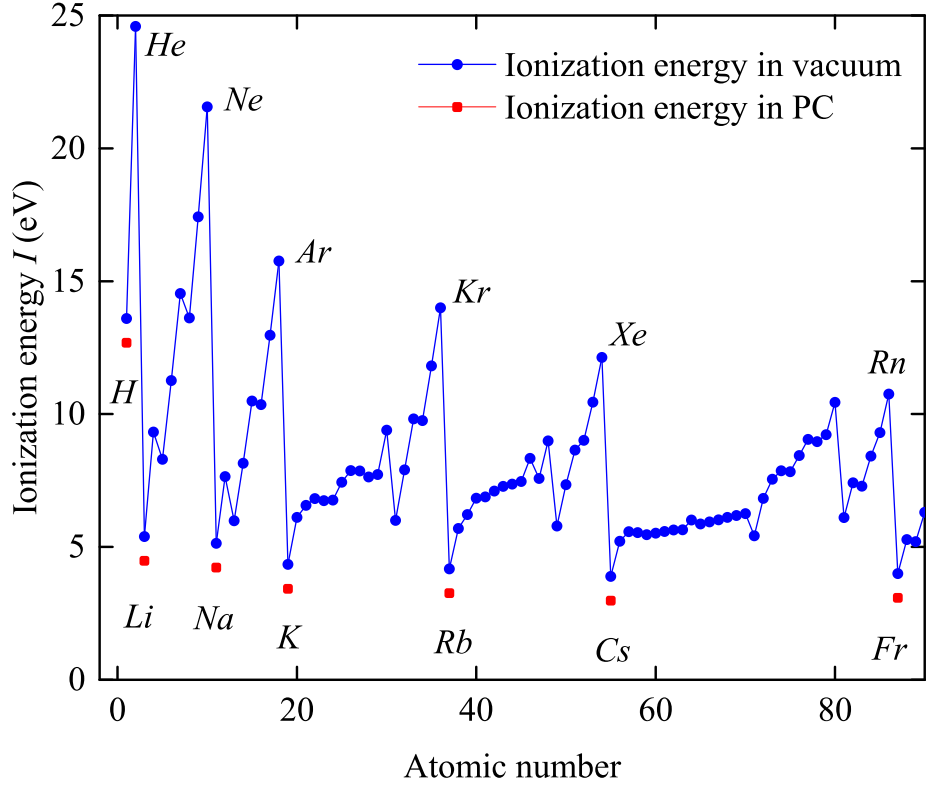}
\caption{\label{fig3} The comparison of ionization energy of hydrogen atom and the alkali metals for the case of vacuum (blue dots) and PC
medium (red squares) on the base of metamaterial with $\textit{a}$ = 30 $nm$, $g$ = 0.7 $nm$. The ionization energy correction has a value
$\delta E_{ion}^{pc}$ = $-$ 0.91 $eV$.}
\end{figure}

\begin{figure}[h]
\centering
\includegraphics[width=0.7\linewidth]{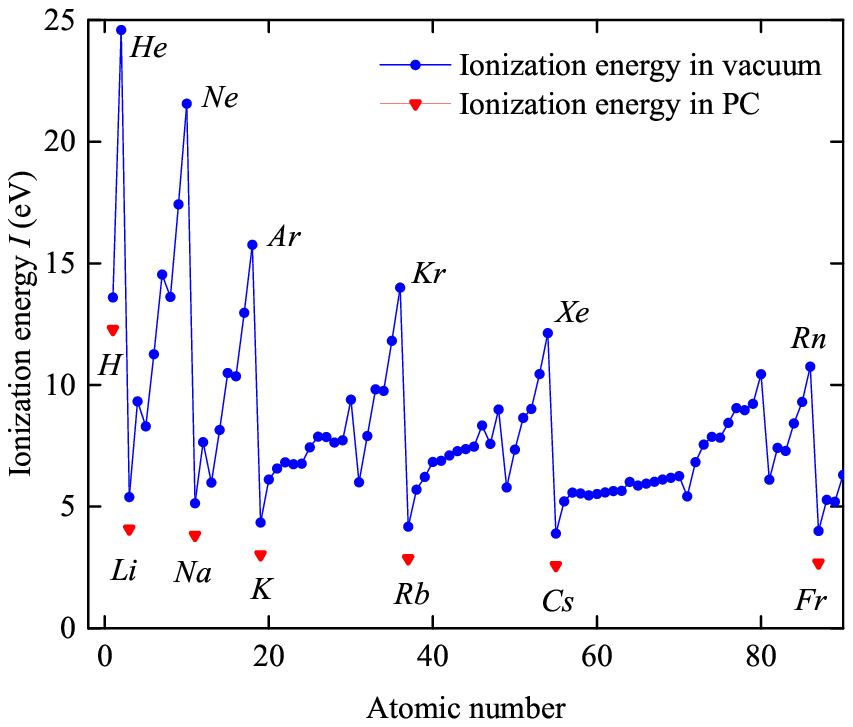}
\caption{\label{fig4} The comparison of ionization energy of hydrogen atom and the alkali metals for the case of vacuum (blue dots) and PC
medium (red triangles) on the base of metamaterial with $\textit{a}$ = 30 $nm$, $g$ = 0.5 $nm$. The ionization energy correction has a value
$\delta E_{ion}^{pc}$ = $-$ 1.32 $eV$.}
\end{figure}

After substituting Eqs.~(\ref{eqn6}) and (\ref{eqn9}) into Eq.~(\ref{eqn13}) the ionization energy correction can be represented in the form:
\begin{equation}\label{eqn14}
\begin{split}
\delta E_{ion}^{pc} = -\frac{2\alpha }{{3\pi }}\sum\limits_{n,\;G}{\left[ {\int\limits {{k_\rho }d{k_\rho }} \int\limits_{FBZ} {d{k_z}} }
\right.} \left( {\frac{{{{\left| {E_{{\bf{k}}n1} \left( G \right)} \right|}^2}}}{{\omega _{{\bf{k}}n {1} }^2}}}
{\left. {\frac{{k_{Gz}^2 - 2k_\rho ^2}}{{k_\rho ^2 + k_{Gz}^2}} + \frac{{{{\left| {E_{{\bf{k}}n2} \left( G \right)}
\right|}^2}}}{{\omega _{{\bf{k}}n{2} }^2}}} \right)} \right].
\end{split}
\end{equation}
When $\textbf{k} \to \infty$ the PC medium is considered as a free space with dielectric constant $\varepsilon(\textbf{r}) = 1$, and it loses
anisotropy \cite{Wang2004giant}. This condition manifests that the integrand of Eq.~(\ref{eqn14}) tends to zero. The absolute value $\delta E_{ion}^{pc}$ increases quadratically with refractive index of the PC's material host, and for this reason, the materials with a highly
tunable refractive index in a wide spectral range of light are needed (See Methods for details of highly tunable refractive index
metameterials). We assume, that metamaterial consisting of $Au$ nanoparticles (AuNPs) ensemble coated with ${HfO_2}$ allows one to achieve the unnaturally highly tunable refractive index (Fig.~\ref{fig2}).
To demonstrate the tunability of the ionization energies of hydrogen atom and the alkali metals we have estimated the ionization energy
corrections $\delta E_{ion}^{pc}$ of these atoms placed in voids of PC on the base of high-index metamaterial with $a$ = 30 $nm$, $g$ = 0.7 $nm$,
and $a$ = 30 $nm$, $g$ = 0.5 $nm$, and are equal to $-$ 0.91 $eV$ and $-$ 1.32 $eV$, respectively. The comparisons of ionization energies for the case of vacuum and the PC medium are represented in Fig.~\ref{fig3} and Fig.~\ref{fig4}.

\section*{Conclusion}

We have shown that the modification of the interaction of an atom placed into the air voids of a PC with its own electromagnetic
field gives rise to the significant change in its ionization energy. The absolute value of the ionization energy correction increases
quadratically with a refractive index of the material host of the PC medium. The effect is strongly enhanced when the PC is made from the
high-index metamaterials, and is tunable. Controlling the geometrical parameters of nanoparticle superlattice of metamaterial can allow one to control chemical reactions that strongly depend on the atomic ionization energies. The effect under study allows one to synthesize novel chemical compounds that could be used in pharmaceutical and other medical-related activities.

\section*{Methods}
\subsection*{The description of the Bloch structure and the self-energy correction ${{\Delta }{E_{em}^{pc}}}(\textbf{p}).$} Because of the spatial modulation of the refractive index the eigenstates of photons in PC differ significantly from those in vacuum or uniform media. Solving quantum-field Maxwell's equations gives rise to the photon states having the Bloch structure similar to that of the
states of electrons in ordinary crystals \cite{Ashcroft1976}. These vectors are the eigenvectors corresponding to the energy $\omega_n$ and
momenta $\bf{k}+\bf{G}$, where $n$ is a band index, the value of $\bf{k}$ is limited by the first Brillouin zone (FBZ) and $\bf{G}$ is the
reciprocal lattice vector of the photonic crystal (${\bf{G}} = N_1 {\bf{b}}_1  + N_2 {\bf{b}}_2 + N_3 {\bf{b}}_3$ with ${\bf{b}}_i$ being
primitive basis vectors of a reciprocal lattice). Because of the translation symmetry the eigenfrequencies of the structure are usually
computed within the first Brillouin zone. By introducing the operators ${\hat a_{{\bf{k}}n}^\dag }$ and ${{{\hat a}_{{\bf{k}}n}}}$ that
describe the creation and annihilation of the photon in the state $\left| {{\bf{k}}n} \right\rangle$, respectively ($\hat a_{{\bf{k}}n}^\dag
\left| 0 \right\rangle  = \left| {{\bf{k}}n} \right\rangle$ and ${{\hat a}_{{\bf{k}}n}}\left| {{\bf{k}}n} \right\rangle  = \left| 0
\right\rangle$), one can construct a modified "free" Hamiltonian ${H_0} = \sum\limits_{{\bf{k}}n} {{\omega _{{\bf{k}}n}}} \hat
a_{{\bf{k}}n}^\dag \hat a_{{\bf{k}}n}^{}$ and quantized vector potential (here and below, we use the natural system of units in which $\hbar
= c = 1$):
\begin{equation}\label{eqn15}
\begin{split}
{\widehat{\bf{A}}_{pc}}({\bf{r}},t) = \sum\limits_{{\bf{k}},\;n} {\left[ {{{\bf{A}}_{{\bf{k}}n}}({\bf{r}}){{\hat a}_{{\bf{k}}n}}{e^{ -
i{\omega _{{\bf{k}}n}}t}} + {\bf{A}}_{{\bf{k}}n}^*({\bf{r}})\hat a_{{\bf{k}}n}^\dag {e^{i{\omega _{{\bf{k}}n}}t}}} \right]},
\end{split}
\end{equation}
where ${{\bf{A}}_{{\bf{k}}n}}({\bf{r}}) = {1 \mathord{\left/{\vphantom {1 {\sqrt {V{\omega _{{\bf{k}}n}}} }}} \right.
 \kern-\nulldelimiterspace} {\sqrt {V{\omega _{{\bf{k}}n}}} }} {{\bf{E}}_{{\bf{k}}n}}({\bf{r}})$ with $\textbf{E}_{\textbf{k}n}(\textbf{r})$
 being the Bloch eigenfunctions satisfying the following orthonormality condition:
\begin{equation}\label{eqn16}
\begin{split}
\int\limits_V {{d^3}r\varepsilon ({\bf{r}}){\bf{E}}_{{\bf{k}}n}^{}({\bf{r}}){\bf{E}}_{{\bf{k}}'n'}^*({\bf{r}})}  = V{\delta
_{{\bf{kk}}'}}{\delta _{nn'}},
\end{split}
\end{equation}
where $\textit{V}$ is the normalization volume of a PC. These eigenfunctions can be expanded as ${\bf{E}}_{{\bf{k}}n} ({\bf{r}}) =
\sum_{\bf{G}}{{\bf{E}}_{{\bf{k}}n} ({\bf{G}})e^{i\left( {{\bf{k}} + {\bf{G}}} \right) \cdot {\bf{r}}} }$, where ${\bf{E}}_{{\bf{k}}n}
({\bf{G}})$ are the Bloch eigenfunctions in the momentum representation.

The PC correction to the electron mass being a result of the modification of the interaction of an electron with the electromagnetic field in
the PC medium is a difference $\Delta E_{em}^{pc}(\textbf{p})$ between the self-energy corrections to kinetic energy in the PC medium and in vacuum
\cite{Gainutdinov2012}
\begin{eqnarray}\label{eqn17}
\begin{split}
{{\Delta }{E_{em}^{pc}}}(\textbf{p}) = {\sum\limits_{{\bf{p}}'}\sum\limits_{{\bf{k}},\;{{n}}} {\frac{{\left\langle {\bf{p}}
\right|H_I^{pc}\left| {{\bf{p}}';{\bf{k}},n} \right\rangle \left\langle {{\bf{p}}';{\bf{k}},n} \right|H_I^{pc}\left| {\bf{p}} \right\rangle
}}{{\frac{{{{\bf{p}}^2}}}{{2{m_e}}} - \frac{{{\bf{p}}{'^2}}}{{2{m_e}}} - {\omega _{{\bf{k}}n}}}}} }
- {\sum\limits_{{\bf{p}}'}\sum\limits_{{\bf{k}},\;{\lambda }} \frac{{\left\langle {\bf{p}} \right|H_I^{}\left|
{{\bf{p}}';{\bf{k}},{\bm{\varepsilon}_\lambda }} \right\rangle \left\langle {{\bf{p}}';{\bf{k}},{\bm{\varepsilon}_\lambda }} \right|H_I^{}\left|
{\bf{p}} \right\rangle }}{{\frac{{{{\bf{p}}^2}}}{{2{m_e}}} - \frac{{{\bf{p}}{'^2}}}{{2{m_e}}} - \left| {\bf{k}} \right|}}}.
\end{split}
\end{eqnarray}
Taking into account that in the nonrelativistic limit the Hamiltonian $H_I$ describing the interaction of the electron with the
electromagnetic field may be written in the form $H_I=-\frac{e}{m_e}{\widehat{\textbf{p}}} \cdot \widehat {\bf{A}}({\bf{r}})$, and the
PC-medium Hamiltonian is modified by replacing $\widehat {\bf{A}}({\bf{r}})$ with $\widehat {\bf{A}}_{pc}({\bf{r}})$, this equation can be
rewritten in the form (see Supplementary Section 1 for details)
\begin{equation}\label{eqn18}
\begin{split}
{{\Delta }{E_{em}^{pc}}}(\textbf{p}) = - \frac{\alpha }{{{2m_e^2\pi ^2}}}\left( \sum\limits_{\lambda } {\int\limits_{}
{\frac{{{d^3}{\bf{k}}}}{{2\left| {\bf{k}} \right|}}\frac{{{{\left| {{\bf{p}} \cdot {\bm{\varepsilon} _\lambda }({\bf{k}})}
\right|}^2}}}{{\frac{{{{\bf{p}}^2}}}{{2{m_e}}} - \frac{{{{\left( {{\bf{p}} - {\bf{k}}} \right)}^2}}}{{2{m_e}}} - \left| {\bf{k}}
\right|}}{\mkern 1mu} } }  - {{\sum\limits_{n,\;{\bf{G}}} {\int\limits_{FBZ} {\frac{{{d^3}{\bf{k}}}}{{{\omega _{{\bf{k}}n}}}}\frac{{{{\left| {{\bf{p}} \cdot {{\bf{E}}_{{\bf{k}}n}}\left( {\bf{G}} \right)} \right|}^2}}}{{\frac{{{{\bf{p}}^2}}}{{2{m_e}}} - \frac{{{{\left( {{\bf{p}} - {\bf{k}} - {\bf{G}}} \right)}^2}}}{{2{m_e}}} - {\omega _{{\bf{k}}n}}}}{\mkern 1mu} } } }} \right) = \\
= - \frac{{\bf{p}}^2}{{{2m_e^2}}}\cdot (\Delta m_{em}^{pc} - \Delta m_{em}) + O{\left( \frac{{\bf{k}}^2}{2m_e} \right)} + O{\left( {\left| {\bf{p}}\right|} \right)^3}.
\end{split}
\end{equation}
The first term on the right-hand part of Eq.~(\ref{eqn18}) is just the ordinary low-energy part of the self-energy of an electron in vacuum
that appears in the second-order perturbation theory \cite{Bjorken1965relativistic,Schweber2011}, and the second one is the modified
self-energy in the PC medium \cite{Gainutdinov2012}. It is important that at the photon energies higher than a few tens of $eV$ the refractive
indices approach 1, and hence the contributions from these terms substitute each other.

\subsection*{Dispersion relations for TE and TM Bloch modes.} $\omega _{{\bf{k}}n1}$ and $\omega _{{\bf{k}}n2}$ are dispersion relations for TE and TM Bloch modes satisfying following transcendent equation \cite{Skorobogatiy2009}
\begin{equation}\label{eqn19}
\begin{split}
 \cos \left( {k_z \left( {d_h  + d_l } \right)} \right) = \cos \left( {k_z^h d_h } \right)\cos \left( {k_z^l d_l } \right)  - {{\left( {r_{ 1, \;2 }  + r_{ 1, \;2 }^{ - 1} } \right)} \mathord{\left/  {\vphantom {{\left( {r_{ 1, \;2 }  + r_{ 1, \;2 }^{ - 1} } \right)} 2}} \right.
 } 2} \cdot \sin \left( {k_z^h d_h } \right)\sin \left( {k_z^l d_l } \right), \\
 \end{split}
 \end{equation}
where ${r_1} = {{k_z^l} \mathord{\left/{\vphantom {{k_z^l} {k_z^h}}} \right.\kern-\nulldelimiterspace} {k_z^h}}, {r_2} = n_h^2{r_1}, k_z^i  =
\sqrt {{{{\omega }}_{{\bf{k}}n}^2} n _i^2  - k_\rho^2 }$
with $d_h$ and $d_l$ being the thicknesses corresponding to the layers of the 1D PC with higher (\textit{h}) and lower (\textit{l}) refractive
index $n_h$ and $n_l = 1$ (air voids), $k_\rho=\sqrt {k_x^2+k_y^2}$.

\subsection*{Highly tunable refractive index metamaterials.} Appropriate refractive properties are demonstrated by optical thin films
formed by nonabsorbing dielectric material such as $HfO_2$ ($n_d$ $\sim$ 2$\div$3 until 10 $eV$) \cite{Franta2015}. At the same time,
it has been recently shown \cite{Lee2015,Chung2016,Kim2016,Kim2018}, that the precise control over the geometrical parameters of
nanoparticle superlattice monolayer leads to the dramatic increase of refractive index far beyond the naturally accessible regime.
According to the theory of the optical effective media \cite{Chung2016}, which is in good agreement with the experiment \cite{Kim2016,Kim2018},
the effective refractive index of such metamaterial is determined by the formula $n_{eff} = [(a/g)\varepsilon_d]^{1/2}$ with array period
$a$, the gap between particles $g$ and the permittivity of the gap-filling dielectric $\varepsilon_d$. We assume, that metamaterial
consisting of $Au$ nanoparticles (AuNPs) ensemble coated with ${HfO_2}$ having $a$ = 30 $nm$, $g$ = 0.7 $nm$ and $a$ = 30 $nm$, $g$ = 0.5 $nm$
allows one to achieve the unnaturally highly tunable refractive index (Fig.~\ref{fig2}).

\section*{Acknowledgements}
The authors are grateful to Dr. A. A. Akhmadeev (Kazan Federal University, Tatarstan Academy of Sciences) for useful discussions.

\section*{Author contributions statement}

R.Kh.G. suggested the idea to control the ionization energies of atoms in the photonic crystal medium, provided quantum electrodynamic analysis of the effect under study, and wrote the main manuscript text. A.I.G. and M.A.Kh. provided analysis and calculation of the effect of the metamaterials on the ionization energies of atoms. M.Kh.S. provided the derivation of finiteness of the PC correction to the electron mass. All authors reviewed the manuscript.

\section*{Additional information}
\noindent $\textbf{Competing Interests:}$ The authors declare no competing interests.

\newpage
\begin{center}
\huge{\bf{Ionization energy of atoms in photonic crystals  (Supplementary Information)}}
\end{center}
\vspace{10mm}

\section*{1. Correction to the self-energy of an electron in the PC medium}
In this section we present a derivation of Eq.~(\ref{eqn18}) of the main text. We start from Eq.~(\ref{eqn17})
\begin{eqnarray}\label{Seqn11}
\begin{split}
{{\Delta }{E_{em}^{pc}}}(\textbf{p}) = {\sum\limits_{{\bf{p}}'}\sum\limits_{{\bf{k}},\;{{n}}} {\frac{{\left\langle {\bf{p}}
\right|H_I^{pc}\left| {{\bf{p}}';{\bf{k}},n} \right\rangle \left\langle {{\bf{p}}';{\bf{k}},n} \right|H_I^{pc}\left| {\bf{p}} \right\rangle
}}{{\frac{{{{\bf{p}}^2}}}{{2{m_e}}} - \frac{{{\bf{p}}{'^2}}}{{2{m_e}}} - {\omega _{{\bf{k}}n}}}}} }
- {\sum\limits_{{\bf{p}}'}\sum\limits_{{\bf{k}},\;{\lambda }} \frac{{\left\langle {\bf{p}} \right|H_I^{}\left|
{{\bf{p}}';{\bf{k}},{\bm{\varepsilon}_\lambda }} \right\rangle \left\langle {{\bf{p}}';{\bf{k}},{\bm{\varepsilon}_\lambda }} \right|H_I^{}\left|
{\bf{p}} \right\rangle }}{{\frac{{{{\bf{p}}^2}}}{{2{m_e}}} - \frac{{{\bf{p}}{'^2}}}{{2{m_e}}} - \left| {\bf{k}} \right|}}}.
\end{split}
\end{eqnarray}
The matrix element $\left\langle {{\bf{p}}';{\bf{k}},n} \right|H_I^{pc}\left| {\bf{p}} \right\rangle $ of the interaction Hamiltonian $H_I^{pc}=-\frac{e}{m_e}{\widehat{\textbf{p}}} \cdot \widehat {\bf{A}}_{pc}({\bf{r}})$ can be represented in the form
\begin{equation}\label{Seqn12}
\left\langle {{\bf{p}}';{\bf{k}},n} \right|H_I^{pc}\left| {\bf{p}} \right\rangle  =  - \frac{e}{m_e}\int {{d^3}r} \Psi _{{\bf{p}}'}^*({\bf{r}})( - i{\nabla _{\bf{r}}}{{\bf{A}}_{{\bf{k}}n}}({\bf{r}}))\Psi _{\bf{p}}^{}({\bf{r}}) = \frac{e}{{m_e{V^{3/2}}\sqrt {{\omega _{{\bf{k}}n}}} }}\int {{d^3}r} {e^{ - i{\bf{p}}'{\bf{r}}}}(i{\nabla _{\bf{r}}}{{\bf{E}}_{{\bf{k}}n}}({\bf{r}})){e^{i{\bf{pr}}}}
\end{equation}
with ${\Psi _{\bf{p}}}({\bf{r}})$ being the normalized wave function of the electron state ${\Psi _{\bf{p}}}({\bf{r}}) = \left\langle {{\bf{r}}}\mathrel{\left | {\vphantom {{\bf{r}} {\bf{p}}}}\right. \kern-\nulldelimiterspace}{{\bf{p}}} \right\rangle$. Here we have taken into account that ${\Psi _{\bf{p}}} = {e^{i{\bf{pr}}}}/\sqrt V $ for ${\bf{r}} \in V$ and ${\Psi _{\bf{p}}} = 0$ for ${\bf{r}} \notin V$. For $\left\langle {{\bf{p}}';{\bf{k}},n} \right|H_I^{pc}\left| {\bf{p}} \right\rangle$ we get
\begin{equation}\label{Seqn13}
\left\langle {{\bf{p}}';{\bf{k}},n} \right|H_I^{pc}\left| {\bf{p}} \right\rangle  =  - \frac{e}{m_e}\frac{1}{{\sqrt {V{\omega _{{\bf{k}}n}}} }}\sum\limits_{\bf{G}} {{\bf{p}} \cdot {{\bf{E}}_{{\bf{k}}n}}({\bf{G}})} {\delta _{{\bf{p}},{\bf{q}}}}
\end{equation}
with ${\bf{q}} = {\bf{p'}} + {\bf{k}} + {\bf{G}}$. In the same way, for $\left\langle {\bf{p}} \right|H_I^{pc}\left| {{\bf{p}}';{\bf{k}},n} \right\rangle$ we find
\begin{equation}\label{Seqn14}
\left\langle {\bf{p}} \right|H_I^{pc}\left| {{\bf{p}}';{\bf{k}},n} \right\rangle  =  - \frac{e}{m_e}\frac{1}{{\sqrt {V{\omega _{{\bf{k}}n}}} }}\sum\limits_{\bf{G}} {{\bf{p}} \cdot {\bf{E}}_{{\bf{k}}n}^ * ({\bf{G}})} {\delta _{{\bf{p}},{\bf{q}}}}.
\end{equation}
Correspondingly, for the matrix elements of the interaction Hamiltonian in the free space, we have
\begin{equation}\label{Seqn15}
\left\langle {{\bf{p}}';{\bf{k}}{\bf{,}}{\bm{\varepsilon}_\lambda }} \right|H_I^{}\left| {\bf{p}} \right\rangle  =  - \frac{e}{m_e}\frac{1}{{\sqrt {2V\left| {\bf{k}} \right|} }}\sum\limits_\lambda  {{\bf{p}} \cdot {\bm{\varepsilon}_\lambda }({\bf{k}})} {\delta _{{\bf{p}},{\bf{q}}}},
\end{equation}
\begin{equation}\label{Seqn16}
\left\langle {{\bf{p}}} \right|H_I^{}\left| {{\bf{p}}';{\bf{k}}{\bf{,}}{\bm{\varepsilon}_\lambda }} \right\rangle  =  - \frac{e}{m_e}\frac{1}{{\sqrt {2V\left| {\bf{k}} \right|} }}\sum\limits_\lambda  {{\bf{p}} \cdot {\bm{\varepsilon}_\lambda }({\bf{k}})} {\delta _{{\bf{p}},{\bf{q}}}}
\end{equation}
with ${\bf{q}} = {\bf{p'}} + {\bf{k}}$. Substituting these matrix elements of interaction Hamiltonians $H_I^{pc}$ and $H_I$ into Eq.~(\ref{Seqn11}), and replacing the discrete sums by integrals $\sum\nolimits_{{\bf{k}}, \;n}  \, \to \tfrac{V}{{{{(2\pi )}^3}}}\sum\nolimits_n  \,\smallint {d^3}{\bf{k}}$ and $\sum\nolimits_{\bf{k}}  \, \to \tfrac{V}{{{{(2\pi )}^3}}}\smallint {d^3}{\bf{k}}$ we get
\begin{equation}\label{Seqn17}
\begin{split}
{{\Delta }{E_{em}^{pc}}}(\textbf{p}) = - \frac{\alpha }{{{2m_e^2\pi ^2}}}\left( \sum\limits_{\lambda } {\int\limits_{}
{\frac{{{d^3}{\bf{k}}}}{{2\left| {\bf{k}} \right|}}\frac{{{{\left| {{\bf{p}} \cdot {\bm{\varepsilon} _\lambda }({\bf{k}})}
\right|}^2}}}{{\frac{{{{\bf{p}}^2}}}{{2{m_e}}} - \frac{{{{\left( {{\bf{p}} - {\bf{k}}} \right)}^2}}}{{2{m_e}}} - \left| {\bf{k}}
\right|}}{\mkern 1mu} } }  - {{\sum\limits_{n,\;{\bf{G}}} {\int\limits_{FBZ} {\frac{{{d^3}{\bf{k}}}}{{{\omega _{{\bf{k}}n}}}}\frac{{{{\left| {{\bf{p}} \cdot {{\bf{E}}_{{\bf{k}}n}}\left( {\bf{G}} \right)} \right|}^2}}}{{\frac{{{{\bf{p}}^2}}}{{2{m_e}}} - \frac{{{{\left( {{\bf{p}} - {\bf{k}} - {\bf{G}}} \right)}^2}}}{{2{m_e}}} - {\omega _{{\bf{k}}n}}}}{\mkern 1mu} } } }} \right).
\end{split}
\end{equation}
\section*{2. Correction to the electron mass in one-dimensional PCs}
The PC correction to the electron mass can be represented in the form
\begin{equation}\label{Seqn21}
\delta {m_{pc}} = \Delta m_{em}^{pc} - \Delta m_{em}^{vac},
\end{equation}
where $\Delta m_{em}^{pc}$ and $\Delta m_{em}^{vac}$ are the total electromagnetic mass in PC and in vacuum. They are both divergent in standard QED, but this PC correction to the electromagnetic mass is finite and could be calculated. The PC correction to the electron mass is given by equation (Eq.~(\ref{eqn6}) in the main text)\cite{Gainutdinov2012}
\begin{equation}\label{eqn211}
\delta m_{pc} = \frac{\alpha }{{\pi ^2 }}\left[ {\sum\limits_{n,\;{\bf{G}}} {\int {{k_\rho }d{k_\rho }} \int\limits_{FBZ}^{} {d{k_z}\int {d\varphi } } }\frac{{\left| {{{\bf{I}}_{\bf{p}}}\cdot {\bf{E}}_{{\bf{k}}n} ({\bf{G}})} \right|^2 }}{{\omega _{{\bf{k}}n}^2}} } -  {\int {\frac{{d^3 {\bf{k}}}}{{2{\bf{k}}^2 }}} \sum\limits_{\lambda  = 1}^2  \,\mathop {\left| { {{\bf{I}}_{\bf{p}}}\cdot
 \bm{\varepsilon} _\lambda  ({\bf{k}})} \right|}\nolimits^2 } \right].
\end{equation}
Representing the electromagnetic field in 1D PC in the Bloch form (Eq.~(\ref{eqn8}) in the main text), the PC correction can be rewritten as
\begin{equation}\label{Seqn22}
\delta m_{pc} = \Delta {m_1} + \Delta {m_2} + \Delta {m_3} - \Delta m_{em}^{vac}
\end{equation}
with
\begin{equation}\label{Seqn23}
\Delta {m_1} = \frac{\alpha }{{{\pi ^2}}}\sum\limits_{n,\ G} {{\int {{k_\rho }d{k_\rho }} \int\limits_{FBZ}^{} {d{k_z}\int {d\varphi } } } {\frac{{{1}}}{{\omega _{{\bf{k}}n1}^2}} {{{\left( {f_{{\bf{p}} \cdot {\bm{\varepsilon}}_{{1}} }^{}E_{{\bf{k}}n1}(G)} \right)}^2}} } },
\end{equation}
\begin{eqnarray}\label{Seqn24}
\begin{split}
\Delta {m_2} = \frac{\alpha }{{{\pi ^2}}}\sum\limits_{n,\;G}^{} {{\int {{k_\rho }d{k_\rho }} \int\limits_{FBZ}^{} {d{k_z}\int {d\varphi } } } {\frac{{2 {{f_{{\bf{p}} \cdot {\bm{\varepsilon}}_{{1}}}}{E_{{\bf{k}}n1}}(G)} }}{{\omega _{{\bf{k}}n1}}}} } \cdot \frac{{ {{f_{{\bf{p}} \cdot {\bm{\varepsilon}}_{{2}}}}{E_{{\bf{k}}n2}}(G)} }}{{\omega _{{\bf{k}}n2}}},
\end{split}
\end{eqnarray}
\begin{equation}\label{Seqn25}
\Delta {m_3} = \frac{\alpha }{{{\pi ^2}}}\sum\limits_{n,\ G} {{\int {{k_\rho }d{k_\rho }} \int\limits_{FBZ}^{} {d{k_z}\int {d\varphi } } }{\frac{1}{{\omega _{{\bf{k}}n2}^2}} {{{\left( {f_{{\bf{p}} \cdot {\bm{\varepsilon}}_{{2}} }^{}E_{{\bf{k}}n2}(G)} \right)}^2}} } },
\end{equation}
where $f_{{\bf{p}} \cdot {\bm{\varepsilon}}}^{} = \sin {\Theta _{\bm{\varepsilon}}}\sin \Theta \cos (\Phi  - {\Phi _{\bm{\varepsilon}}}) + \cos {\Theta _{\bm{\varepsilon}}}\cos \Theta$ is the scalar product between the unit vector of electron's momentum with angular coordinates $\left( {\Theta ,\Phi } \right)$ and ${\bm{\varepsilon}}_{{1}}(\bf{k_G}) $ and ${\bm{\varepsilon}}_{{2}}(\bf{k_G})$ are the field unit vectors with angular coordinates $\left( {{\Theta _{\bm{\varepsilon}}},{\Phi _{\bm{\varepsilon}}}} \right)$ and ${\bf{k_G}} = {\bf{k}} + G{{\bf{e}}_z}$ in the case of 1D PC. Taking the integrals in Eqs.~(\ref{Seqn23}--\ref{Seqn25}) over the azimuthal angle $\varphi_k$ in the cylindrical coordinate system, we get
\begin{eqnarray}\label{Seqn26}
\begin{split}
\Delta {m_1} = \frac{{{\alpha }}}{{{\pi }}}\sum\limits_{n,\ G}^{} {\int {{k_\rho }d{k_\rho }} \int\limits_{FBZ}^{} {d{k_z}\frac{{{{\left| {{E_{{\bf{k}}n1}}(G)} \right|}^2}}}{{{{\omega }}_{{\bf{k}}n1}^2}} \cdot } }\left( {\frac{{(k_z + G)^2}}{{k_\rho ^2 + (k_z + G)^2}}{{\sin }^2}\Theta  + 2\frac{{k_\rho ^2}}{{k_\rho ^2 + (k_z + G)^2}}{{\cos }^2}\Theta } \right),
\end{split}
\end{eqnarray}
\begin{equation}\label{Seqn27}
\Delta {m_2}=0,
\end{equation}
\begin{equation}\label{Seqn28}
\Delta {m_3} = \frac{{{\alpha }}}{{{\pi }}}\sum\limits_{n,\ G}^{} {\int {{k_\rho }d{k_\rho }} \int\limits_{FBZ}^{} {d{k_z}} \frac{{{{\left| {{E_{{\bf{k}}n2}}(G)} \right|}^2}}}{{{{\omega }}_{{\bf{k}}n2}^2}}{{\sin }^2}\Theta },
\end{equation}
where $k_\rho, k_z + G$ are radial component and $z$-component of the wave vector $\bf{k}$ in cylindrical coordinate system. The electromagnetic mass of an electron in vacuum takes the form $\frac{{4{{\alpha }}}}{{3{{\pi }}}}{\int {dk}}$. Then the PC correction to the electron mass Eq.~(\ref{Seqn22}) in 1D PC is equal:
\begin{equation}\label{Seqn29}
\begin{split}
\delta {m_{pc}} = \frac{\alpha }{\pi }\sum\limits_{n,\;G}^{} {\int {{k_\rho }d{k_\rho }} \int\limits_{FBZ}^{} {d{k_z}\left[ {\frac{{{{\left| {{E_{{\bf{k}}n1}}(G)} \right|}^2}}}{{\omega _{{\bf{k}}n1}^2}}\frac{\left( {{{({k_z} +G)}^2}{{\sin }^2}\Theta  + 2k_\rho ^2{{\cos }^2}\Theta } \right)}{{k_\rho ^2 + {{({k_z} + G)}^2}}}} +  \frac{{{{\left| {{E_{{\bf{k}}n2}}(G)} \right|}^2}}}{{\omega _{{\bf{k}}n2}^2}} \cdot {{\sin }^2}\Theta \right]}} - \frac{{4\alpha }}{{3\pi }}\int {dk}.
\end{split}
\end{equation}
The operator of the observable $\delta m_{pc}$ can be presented in the form
\begin{equation}\label{Seqn210}
\widehat {\delta {m}}_{pc} = A + \left( {{{\widehat {\bf{I}}}_{\bf{p}}} \cdot {{\widehat {\bf{I}}}_{pc}}} \right)^{2}B,
\end{equation}
where ${\widehat {\bf{I}}_{\bf{{p}}}} = \frac{\widehat{\bf{p}}}{\left| {\bf{p}} \right|}$  being the direction of the electron momentum,  ${{{\widehat {\bf{I}}}_{pc}}}$ is the unit vector of the 1D PC crystal axis that coincides with vector ${\bf{e}}_z$, $\left( {{{\widehat {\bf{I}}}_{\bf{p}}} \cdot {{\widehat {\bf{I}}}_{pc}}} \right) = \cos \Theta$, and
\begin{equation}
A = \frac{{{\alpha }}}{{{\pi }}}\sum\limits_{n,\;G}^{} {\int {{k_\rho }d{k_\rho }} \int\limits_{FBZ}^{} {d{k_z}\left( {\frac{{{{\left| {E_{{\bf{k}}n1}^{}\left( G \right)} \right|}^2}}}{{{{\omega }}_{{\bf{k}}n1}^2}}\frac{{(k_z + G)^2}}{{k_\rho ^2 +(k_z + G)^2}}}\right.}}
\left. { + \frac{{{{\left| {E_{{\bf{k}}n2}^{}\left( G \right)} \right|}^2}}}{{{{\omega }}_{{\bf{k}}n2}^2}}} \right)-\frac{{4{{\alpha }}}}{{3{{\pi }}}}{\int {dk}},\notag \\
\end{equation}
\begin{equation}
B = \frac{{{\alpha }}}{{{\pi }}}\sum\limits_{n,\;G}^{} {\int {{k_\rho }d{k_\rho }} \int\limits_{FBZ}^{} {d{k_z}\left( {\frac{{{{\left| {E_{{\bf{k}}n1}^{}\left( G \right)} \right|}^2}}}{{{{\omega }}_{{\bf{k}}n1}^2}}\frac{{2k_\rho ^2 - (k_z + G)^2}}{{k_\rho ^2 + (k_z + G)^2}} - } \right.} }
\left. {\frac{{{{\left| {E_{{\bf{k}}n2}^{}\left( G \right)} \right|}^2}}}{{{{\omega }}_{{\bf{k}}n2}^2}}} \right).\notag \\
\end{equation}

\section*{3. Finiteness of the PC correction to the electron mass}
In the investigation of the convergence of the integrals in Eq.~(\ref{Seqn29}) it is natural to represent the PC correction to the electromagnetic mass as the sum of the low-energy (LE) and high-energy (HE) parts \cite{Schweber2011}
\begin{equation}\label{Seqn31}
\delta m_{pc} = \delta m_{pc}^{LE} + \delta m_{pc}^{HE},
\end{equation}
where
\begin{equation}\label{Seqn32}
\begin{split}
\delta m_{pc}^{LE} = \frac{\alpha }{{{\pi ^2}}}\left[ {\mathop {\sum\limits_{n,\;G} {\int\limits_{} {{k_\rho }d{k_\rho }} } }\limits^{{\omega _{{\bf{k}}n}} < \Lambda } \int\limits_{FBZ}^{} {d{k_z}\int\limits_0^{2\pi } {d{k_\varphi }}  } } \right.
\sum\limits_{\lambda  = 1}^2 {\frac{{{{\left| {{E_{{\bf{k}}n\lambda }}(G)} \right|}^2}}}{{\omega _{{\bf{k}}n\lambda }^2}}} {\left| {{{\bf{I}}_{\bf{p}}} \cdot {\bm{\varepsilon} _\lambda }({{\bf{k}}_{\bf{G}}})} \right|^2} - \left. {\int\limits_{\left| {\bf{k}} \right| < \Lambda }^{} {\frac{{{d^3}{\bf{k}}}}{{3{{\bf{k}}^2}}}} } \right],
\end{split}
\end{equation}
\begin{equation}\label{Seqn33}
\begin{split}
\delta m_{pc}^{HE} = \frac{\alpha }{{{\pi ^2}}}\left[ {\mathop {\sum\limits_{n,\;G} {\int\limits_{} {{k_\rho }d{k_\rho }} } }\limits^{{\omega _{{\bf{k}}n}} > \Lambda } \int\limits_{FBZ}^{} {d{k_z}\int\limits_0^{2\pi } {d{k_\varphi }}  } } \right.
\sum\limits_{\lambda  = 1}^2 {\frac{{{{\left| {{E_{{\bf{k}}n\lambda }}(G)} \right|}^2}}}{{\omega _{{\bf{k}}n\lambda }^2}}} {\left| {{{\bf{I}}_{\bf{p}}} \cdot {\bm{\varepsilon} _\lambda }({{\bf{k}}_{\bf{G}}})} \right|^2} - \left. {\int\limits_{\left| {\bf{k}} \right| > \Lambda }^{\infty} {\frac{{{d^3}{\bf{k}}}}{{3{{\bf{k}}^2}}}} } \right].
\end{split}
\end{equation}
Here the cutoff $\Lambda$ should be chosen to be much larger than the width $|{{\bf{b}}_z}|$ of FBZ. Obviously, the divergence problem can appear only in the HE part (\ref{Seqn33}). In the limit $\bf{k} \to \infty$ the 1D PC medium is considered as a free space with effective refractive index ${\tilde{n}}_{eff}({\bf{k}})$ defined by equation \cite{Skorobogatiy2009}
\begin{equation}\label{Seqn34}
{\tilde{n}}_{eff}({\bf{k}}) = ({n_h({\bf{k}})d_h + n_l d_l})/({d_h + d_l}),
\end{equation}
where $d_h$ and $d_l$ are the thicknesses corresponding to the layers of the 1D PC with higher (\textit{h}) and lower (\textit{l}) refractive index $n_h({\bf{k}})$ and $n_l = 1$ (air voids). In this limit it is more convenient to make use the extended zone scheme, where the summation on the band index $n$, and the integration in the FBZ are transformed into the integral over all wave vectors in space satisfying the condition $|{\bf{k}}| > \Lambda$
\begin{equation}\label{Seqn35}
\begin{split}
\delta m_{pc}^{HE} = \frac{\alpha }{{{\pi ^2}}}\left[ {\int\limits_{\left| {\bf{k}} \right| > \Lambda }^{\infty} {{k_\rho }d{k_\rho }} \int\limits_{\left| {\bf{k}} \right| > \Lambda }^{\infty} {d{k_z}\int\limits_0^{2\pi } {d{k_\varphi }}  } } \right.
\left. { \sum\limits_{\lambda  = 1}^2 {\frac{{{{\left| {{E_\lambda }({\bf{k}})} \right|}^2}}}{\omega_{\lambda}^{2} ({\bf{k}})}{{\left| {{{\bf{I}}_{\bf{p}}} \cdot {\bm{\varepsilon} _\lambda }({{\bf{k}}})} \right|}^2}}  - \int\limits_{\left| {\bf{k}} \right| > \Lambda }^{\infty} {\frac{{{d^3}{\bf{k}}}}{{3{{\bf{k}}^2}}}} } \right].
\end{split}
\end{equation}
Then dispersion relations $\omega_{\lambda} ({\bf{k}})$ are symmetrical quasi-continuous functions of ${\bf{k}}$ everywhere in reciprocal space \cite{Ashcroft1976}
\begin{equation}\label{Seqn36}
\omega_{\lambda} ({\bf{k}})=\frac{|{\bf{k}}|}{{\tilde{n}}_{eff}({\bf{k}})} \left(1 + \frac{{{\tilde{n}}_{eff}({\bf{k}})}\Delta \omega_{\lambda}({\bf{k}})}{|{\bf{k}}|} \right) \\
\end{equation}
with $\Delta \omega_{\lambda}({\bf{k}})$ being the PC correction to the dispersion relation $\frac{|{\bf{k}}|}{{\tilde{n}}_{eff}({\bf{k}})}$ in isotropic medium and
\begin{equation}\label{Seqn37}
|E_{\lambda}({\bf{k}})|^2 = \frac{1}{2} + \Delta E_{\lambda}({\bf{k}}). \\
\end{equation}
These eigenfunctions of Maxwell's equations with corresponding eigenvalues $\omega_{\lambda} ({\bf{k}})$ are equal to $\frac{1}{2}$ because $\Delta E_{\lambda}({\bf{k}})$ is proportional to $\Delta \omega_{\lambda}({\bf{k}})$.
The correction $\Delta \omega_{\lambda}({\bf{k}})$ is limited by the function \cite{Skorobogatiy2009}
\begin{equation}\label{Seqn38}
\Delta \omega_{\lambda}({\bf{k}}) \leq \frac{n_h({\bf{k}}) - 1}{n_h({\bf{k}}) + 1}\omega. \\
\end{equation}
According to Sellmeier equation \cite{Ghosh1997}, ${{n}}_{h} ({\textbf{k}})$ can be represented as a power series
\begin{equation}\label{Seqnnh39}
{{n}}_{h} ({\textbf{k}}) = 1 + \frac{C_1}{{\bf{k}}^2} + \frac{C_2}{{\bf{k}}^4} + ..., \\
\end{equation}
where $C_{1,2}$ are some experimentally determined parameters. Under the condition of high-energy photons propagating in 1D PC the corrections $\Delta \omega_{\lambda}({\bf{k}})$ in dispersion relations are vanished because $n_h({\bf{k}}) - 1\rightarrow 0$ as $C_1/{\textbf{k}}^2$ and eigenfrequencies $\omega_{\lambda}({\bf{k}})$ tend to each other. Then the HE energy contribution to the electromagnetic mass is represented in the form
\begin{equation}
\label{Seqn310}
\delta m_{pc}^{HE} = \frac{\alpha }{{3{\pi ^2}}}\int\limits_{|{\bf{k}}| > \Lambda }^\infty  {{d^3}{\bf{k}}{\frac{{\tilde n_{eff}^2({\bf{k}}) - 1}}{{{{\bf{k}}^2}}}
+ O\left( {\frac{{C_1}}{{{{\Lambda}}}}} \right) + O\left( {\frac{{{\bf{b}}_z^2}}{{{{\Lambda}}}}} \right)}}.
\end{equation}
From Eqs.~(\ref{Seqn34}) and (\ref{Seqnnh39}) it follows that
\begin{equation}\label{Seqn311}
\delta {m_{pc}^{HE}} = \frac{\alpha }{{6{\pi ^2}}}\int\limits_{|{\bf{k}}| > \Lambda }^\infty
{\frac{{{d^3}{\bf{k}}}}{{{{\bf{k}}^4}}} { \frac{C_1 d_h}{d_h + d_l} + O\left( {\frac{{C_1}}{{{{\Lambda}}}}} \right) +
O\left( {\frac{{{\bf{b}}_z^2}}{{{{\Lambda}}}}} \right)} }.
\end{equation}
Thus the PC correction to the electromagnetic mass is free from the ultraviolet divergences and hence is an observable.


\begin{thebibliography}{41}
\bibitem{IYPT2019} International Union of Pure and Applied Chemistry (IUPAC), The International Year of the Periodic Table
https://www.iypt2019.org/ (2018).

\bibitem{Yablonovitch1987} Yablonovitch, E. Inhibited spontaneous emission in solid-state physics and electronics. \textit{Phys. Rev. Lett.}
    \textbf{58,} 2059--2062 (1987).

\bibitem{John1987} John, S. Strong localization of photons in certain disordered dielectric superlattices. \textit{Phys. Rev. Lett.}
    \textbf{58,} 2486--2489 (1987).

\bibitem{Lopez2003} Lopez, C. Materials aspects of photonic crystals. \textit{Adv. Mater.} \textbf{15,} 1679--1704 (2003).

\bibitem{Joannopoulos2008} Joannopoulos, J. D., Johnson, S. G., Winn, J. N. \& Meade, R. D. 2nd ed. \textit{Photonic Crystals: Molding the Flow of Light}. (Princeton Univ. Press, 2008).

\bibitem{Soukoulis2001photonic} Soukoulis, C. M. \textit{Photonic Crystals and Light Localization in the 21st Century.} Vol. 563 (Kluwer
    Academic Publishers, 2001).

\bibitem{Von2013bottom} Freymann, G., Kitaev, V., Lotsch, B. V. \& Ozin, G. A. Bottom-up assembly of photonic crystals. \textit{Chem. Soc.
    Rev.} \textbf{42,} 2528--2554 (2013).

\bibitem{John1990quantum} John, S. \& Wang, J. Quantum electrodynamics near a photonic band gap: Photon bound states and dressed atoms.
    \textit{Phys. Rev. Lett.} \textbf{64,} 2418--2421 (1990).

\bibitem{John1991} John, S. \& Wang, J. Quantum optics of localized light in a photonic band gap. \textit{Phys. Rev. B.} \textbf{43,}
    12772--12789 (1991).

\bibitem{Quang1997coherent} Quang, T., Woldeyohannes, M., John, S. \& Agarwal, G. S. Coherent control of spontaneous emission near a photonic
    band edge: a single-atom optical memory device. \textit{Phys. Rev. Lett.} \textbf{79,} 5238--5241 (1997).

\bibitem{Zhu1997quantum} Zhu, S.-Y., Chen, H. \& Huang, H. Quantum interference effects in spontaneous emission from an atom embedded in a
    photonic band gap structure. \textit{Phys. Rev. Lett.} \textbf{79,} 205--208 (1997).

\bibitem{Bay1997atom} Bay, S., Lambropoulos, P. \& M{\o}lmer, K. Atom--atom interaction in strongly modified reservoirs. \textit{Phys. Rev.
    A.} \textbf{55,} 1485--1496 (1997).

\bibitem{Busch2000radiating} Busch, K., Vats, N., John, S. \& Sanders, B. C. Radiating dipoles in photonic crystals. \textit{Phys. Rev. E.}
    \textbf{62,} 4251--4260 (2000).

\bibitem{Aguirre2010tunable} Aguirre, C. I., Reguera, E. \& Stein, A. Tunable colors in opals and inverse opal photonic crystals. \textit{Adv.
    Funct. Mater.} \textbf{20,} 2565--2578 (2010).

\bibitem{Huang2011dirac} Huang, X., Lai, Y., Hang, Z. H., Zheng, H. \& Chan, C. Dirac cones induced by accidental degeneracy in photonic
    crystals and zero-refractive-index materials. \textit{Nat. Mater.} \textbf{10,} 582--586 (2011).

\bibitem{Fenzl2014photonic} Fenzl, C., Hirsch, T. \& Wolfbeis, O. S. Photonic crystals for chemical sensing and biosensing. \textit{Angew.
    Chem.} \textbf{53,} 3318--3335 (2014).

\bibitem{Goban2014atom} Goban, A. et al. Atom--light interactions in photonic crystals. \textit{Nat. Commun.} \textbf{5,} 3808 (2014).

\bibitem{Segal2015controlling} Segal, N., Keren-Zur, S., Hendler, N. \& Ellenbogen, T. Controlling light with metamaterial-based nonlinear
    photonic crystals. \textit{Nat. Photonics} \textbf{9,} 180--184 (2015).

\bibitem{Gainutdinov2012} Gainutdinov, R. K., Khamadeev, M. A. \& Salakhov, M. K. Electron rest mass and energy levels of atoms in the
    photonic crystal medium. \textit{Phys. Rev. A.} \textbf{85,} 053836 (2012).

\bibitem{Zhu2012highly} Zhu, Y. et al. Highly modified spontaneous emissions in $YVO_4$: $Eu^{3+}$ inverse opal and refractive index sensing
application. \textit{Appl. Phys. Lett.} \textbf{100,} 081104 (2012).

\bibitem{Zhu2012inhibited} Zhu, Y. et al. Inhibited long-scale energy transfer in dysprosium doped yttrium vanadate inverse opal.
\textit{J. Phys. Chem. C.} \textbf{116,} 2297--2302 (2012).

\bibitem{Liu2010observation} Liu, Q. et al. Observation of Lamb shift and modified spontaneous emission dynamics
in the $YBO_3: Eu^{3+}$ inverse opal. \textit{Opt. Lett.} \textbf{35,} 2898--2900 (2010).

\bibitem{Roy2010coherent} Roy, C. Coherent control of a three-level atom in a photonic crystal. \textit{J. Phys. B.} \textbf{43,}
235502 (2010).

\bibitem{Yoshie2004vacuum} Yoshie, T. Vacuum Rabi splitting with a single quantum dot in a photonic crystal nanocavity. \textit{Nature}
\textbf{432,} 200 (2004).

\bibitem{Vats2002theory} Vats, N., John, S. \& Busch, K. Theory of fluorescence in photonic crystals. \textit{Phys. Rev. B.} \textbf{65,}
043808 (2002).

\bibitem{Li2001quantum} Li, Z. \& Xia, Y. Quantum optics of localized light in a photonic band gap. \textit{Phys. Rev. B.} \textbf{63,} 121305 (2001).

\bibitem{Wang2004giant} Wang, X.-H., Kivshar, Y. S. \& Gu, B.-Y. Giant lamb shift in photonic crystals. \textit{Phys. Rev. Lett.} \textbf{93,}    073901 (2004).

\bibitem{Wang2005spontaneous} Wang, X.-H., Gu, B.-Y. \& Kivshar, Y. Spontaneous emission and lame shift in photonic crystals. \textit{Sci. Technol. Adv. Mater.}
\textbf{6,} 814--822 (2005).

\bibitem{Gainutdinov2018} Gainutdinov, R., Khamadeev, M., Akhmadeev, A. \& Salakhov, M. \textit{Modification of the Electromagnetic Field in the Photonic Crystal Medium and New Ways of Applying the Photonic Band Gap Materials}. [Vakhrushev, A. (ed.)] \textit{Theoretical Foundations and Application of Photonic Crystals}. Chap. 1., 3--20. (InTech, 2018).

\bibitem{Weinberg1995} Weinberg, S. \textit{The Quantum Theory of Fields: Foundations}. Chap. 14., Vol. 1. (Cambridge Univ. Press, 2005).

\bibitem{Lee2015} Lee, S. Colloidal superlattices for unnaturally high-index metamaterials at broadband optical frequencies. \textit{Opt.
    Express} \textbf{23,} 28170--28181 (2015).

\bibitem{Chung2016} Chung, K., Kim, R., Chang, T. \& Shin, J. Optical effective media with independent control of permittivity and
    permeability based on conductive particles. \textit{Appl. Phys. Lett.} \textbf{109,} 021114 (2016).

\bibitem{Kim2016} Kim, J. Y. et al. Highly tunable refractive index visible-light metasurface from block copolymer self-assembly. \textit{Nat. Commun.} \textbf{7,} 12911 (2016).

\bibitem{Kim2018} Kim, R. et al. Metal nanoparticle array as a tunable refractive index material over broad visible and infrared wavelengths.
    \textit{ACS Photonics} \textbf{5,} 1188 (2018).

\bibitem{Cohen1998atom} Cohen-Tannoudji, C., Dupont-Roc, J. \& Grynberg, G. \textit{The dressed atom approach}. \textit{Atom-photon interactions: basic processes and applications}. Chap. 6., 537--548. (Wiley-VCH,
    1998).

\bibitem{Feynman1948} Feynman, R. P. Space-time approach to non-relativistic quantum mechanics. \textit{Rev. Mod. Phys.} \textbf{20,} 367-–387 (1948).

\bibitem{Skorobogatiy2009} Skorobogatiy, M. \& Yang, J. \textit{Fundamentals of Photonic Crystal Guiding}. (Cambridge Univ. Press,
    2009).

\bibitem{Ashcroft1976} Ashcroft, N. W. \& Mermin, N. D. \textit{Solid State Physics}. (Harcourt College Publishers, 1976).

\bibitem{Bjorken1965relativistic} Bjorken, J. D. \& Drell, S. D. \textit{Relativistic Quantum Mechanics}. (McGraw-Hill, 1965).

\bibitem{Schweber2011} Schweber, S. S. \textit{An Introduction to Relativistic Quantum Field Theory}. (Courier, 2011).

\bibitem{Franta2015} Franta, D., Necas, D. \& Ohlidal, I. Universal dispersion model for characterization of optical thin films over a wide
    spectral range: application to hafnia. \textit{Appl. Opt.} \textbf{54,} 9108--9119 (2015).

\end{thebibliography}

\begin{thebibliography}{5}
\bibitem{Gainutdinov2012} Gainutdinov, R. K., Khamadeev, M. A. \& Salakhov, M. K. Electron rest mass and energy levels of atoms in the
    photonic crystal medium. \textit{Phys. Rev. A.} \textbf{85,} 053836 (2012).

\bibitem{Schweber2011} Schweber, S. S. \textit{An Introduction to Relativistic Quantum Field Theory} (Courier, New York, 2011).

\bibitem{Skorobogatiy2009} Skorobogatiy, M. \& Yang, J. \textit{Fundamentals of Photonic Crystal Guiding} (Cambridge Univ. Press, New York, 2009).

\bibitem{Ashcroft1976} Ashcroft, N. W. \& Mermin, N. D. \textit{Solid State Physics} (Harcourt College Publishers, Harcourt, 1976).

\bibitem{Ghosh1997} Ghosh, G. Sellmeier coefficients and dispersion of thermo-optic coefficients for some optical glasses. \textit{Appl. Opt.} \textbf{36,} 1540--1546 (1997).

\end{thebibliography}
\end{document}